# "Wall Slip Effects on Parameter Estimation for Integral-Type Viscoelastic Models: Insights from High Density Polyethylene and Thermoplastic Elastomer Melts"

**Halil Gevgilili and Dilhan M. Kalyon***

Stevens Institute of Technology

Hoboken, NJ 07030

* dkalyon@stevens.edu

**Synopsis:**

This study explores the impact of wall slip on parameter estimation for integral-type non-linear viscoelastic models with time-strain separable memory functions, specifically examining high-density polyethylene (HDPE) and thermoplastic elastomer (TPE) melts. Significant wall slip was observed in HDPE under non-linear conditions, complicating the characterization of its damping function using step strain data. In contrast, TPE exhibited negligible wall slip across a wide range of strains and shear rates, facilitating successful parameterization of its damping function using the Wagner-type K-BKZ equation and enabling accurate predictions of various material functions. For HDPE, the study investigated alternative approaches to parameterize the K-BKZ equation, including the use of steady and time-dependent material functions at low deformation rates where wall slip effects are minimal. The findings underscore the challenges of estimating parameters in the presence of wall slip and propose strategies to address these challenges, particularly for materials like HDPE, where pronounced slip complicates traditional rheological characterization.

# 1 Background and Introduction:

It is generally recognized that some polymer melts, especially linear polymers, polymeric gels and suspensions exhibit wall slip during rheological characterization and various processing flows. In capillary flow, such wall slip is inferred from discontinuities in the flow curves [[Kalika and Denn (1987); He et al. (2019); Quan and Kalyon (2022)], from the dependence of the flow curves on the surface to volume ratio of the die used [Ramamurthy (1986); Yilmazer and Kalyon (1989); Chen et al. (1993), Kalyon (2005)] or observed directly from the velocity distributions [Migler et al. (1993); Münstedt et al. (2000); Aktas et al. (2014); Ortega-Avila et al. (2016); Medina-Bañuelos et al. (2017)]. The wall slip behavior of various complex fluids including polymer melts in other flows, including steady torsional [Awati et al. (2000)], plane Couette [Hatzikiriakos and Dealy (1991)], step-strain [Gevgilili and Kalyon (2001)], squeeze flow [Kalyon and Tang (2007); Tang and Kalyon (2004)], axial annular flow [Kalyon and Malik (2012)], coextrusion flows [Kalyon (2010)], continuous shear roll milling [Kalyon et al. (2004)] and injection molding flows [Rielly and Price (1961)] are also documented. The wall slip behavior of the polymer melt can be dependent on the materials of construction and on the roughness of the internal die surface [Ramamurthy (1986); Chen et al. (1993)]. Wall slip has significant effects during processing in various geometries [Malik et al. (2014)], including effects on the development of flow instabilities during extrusion [Kalyon and Gevgilili (2003); Birinci and Kalyon (2006); Tang and Kalyon (2008a and 2008b)].

Although wall slip during rheological characterization is so prevalent, the conventional procedures for the estimation of the parameters of various constitutive equations are generally not altered to accommodate wall slip. Generally, the rheometer surfaces are roughened to attempt to eliminate wall slip, leading to the fracturing of the sample and faulty experimentation [Kalyon (2005a)] suggesting that the material functions be determined with smooth surfaces following wall slip analysis [Kalyon (2005b); He et al. (2019); Quan and Kalyon (2022)]. Here, this issue is further probed by using two melts, which are very different in terms of their wall slip behavior. First, a general background on the methods used in fitting the parameters of integral-type constitutive equations, based on the time and strain separability of the memory function, is provided. Second, the material functions in shear and extension of the two melts, one of which exhibits strong wall slip and the other does not, are discussed. Third, the inverse problem was solved to determine the parameters of the damping function by using different material functions and the predictions compared. Finally, our earlier observation that the widely-reported overshoots in shear stress and the first normal stress difference during simple shear could be artifacts of wall slip [Kalyon and Gevgilili (2003)] is elucidated further by probing the sensitivity of the predictions of stress overshoots on the parameters of the constitutive equation.

**Background**

Various non-linear viscoelastic constitutive equations are proposed to relate the stress tensor to the strain and strain rate tensors [Bird et al. (1987); Larson (1999); Graessley (2004); Dealy et al. (2018); Sengul (2021)]. The following nonlinear viscoelastic equation, proposed by Kaye (1962) and Bernstein, Kearsley and Zapas (1963), K-BKZ equation, is widely used,

$$\boldsymbol{\tau}(t) = \int_{-\infty}^{t} \left[ \frac{\partial V(t-t',I_1,I_2)}{\partial I_1} \boldsymbol{\gamma}_{[0]} + \frac{\partial V(t-t',I_1,I_2)}{\partial I_2} \boldsymbol{\gamma}^{[0]} \right] dt' \qquad (1)$$

where $V(t-t',I_1,I_2)$ is a time-dependent elastic energy kernel function, $\boldsymbol{\gamma}_{[0]}$ and $\boldsymbol{\gamma}^{[0]}$ are the relative (finite) strain tensors [Bird et al. (1987)], and $I_1$ and $I_2$ are the first and second invariants of the Finger strain tensor, indicating that the stress at the present time, t depends on the history of the strain for all past times $-\infty<t'<t$. The energy kernel is usually assumed to be a product of time-dependent and strain dependent factors [White et al. (1967)], i.e., $V(t-t',I_1,I_2)=M(t-t')W(I_1,I_2)$. W is the potential function, $M(t-t')=\partial G(t-t')/\partial t'$ is the time-dependent linear viscoelastic memory function and $G(t-t')$ is the relaxation modulus.

A special form of the K-BKZ equation is given by Wagner (1976, 1979)],

$$\boldsymbol{\tau}(t) = \int_{-\infty}^{t} M(t-t')h(I_1,I_2)\boldsymbol{B}(t')dt' \qquad (2a)$$

where $h(I_1,I_2)$ is the damping function, dependent on the first and second invariants, $I_1$ and $I_2$, of the Finger strain tensor, $\boldsymbol{B}(t')$. Thus, the fading memory of the fluid is represented as the product of two terms, one dependent on time alone and the second on strain alone, hence time and strain separability. If the damping function, h, is set to one and the Finger strain tensor, B, is replaced by the relative finite strain tensor $\gamma_{[0]}$, one arrives at the Lodge rubberlike liquid [Lodge 1964].

A widely-used choice for the linear viscoelastic memory function, M(t-t′), is the generalized Maxwell model given as:

$$M(t-t) = \left\{ \sum_{k=1}^{N} \frac{\eta_k}{\lambda_k^2} e^{-(t-t')/\lambda_k} \right\} \qquad (2b)$$

The linear viscoelastic memory function contains 2 sets of parameters, i.e., the spectrum of empirically determined relaxation times, $\lambda_k$, and corresponding viscosities, $\eta_k$, of the individual Maxwell elements. Various experimental approaches for the determination of the parameters of the relaxation spectra were outlined by He et al. (2004).

For simple shear flow, the damping function for shear, $h_s$, can be described as [(Osaki (1976)],

$$h_s(\gamma) = f \exp(-n_1\gamma) + (1-f) \exp(-n_2\gamma) \qquad (3)$$

where f, $n_1$, $n_2$ are material parameters. This equation is referred to as the "double exponential" damping function [Laun (1978)]. For uniaxial extensional deformation, Wagner proposed a different damping function, $h_e$,

$$1 / h_e(\varepsilon) = a' \exp[2\varepsilon] + (1 - a')\exp[m\varepsilon] \qquad (4)$$

where m and a' are empirical constants [Wagner (1978)] and were determined to be m=0.3 and a'=exp[-2$\varepsilon_o$] with $\varepsilon_o$ =3.0, for low density polyethylene. These damping function equations for shear and extension can be combined as [Wagner (1979)],

$$h(I_1, I_2) = f \exp[-n_1 \sqrt{\alpha I_1 + (1-\alpha)I_2}] + (1-f)\exp[-n_2 \sqrt{\alpha I_1 + (1-\alpha)I_2}] \qquad (5)$$

For simple shear, $I_1=I_2$ and Equation 5 reduces to Equation 3. For uniaxial extensional flow $I_1 = \exp(2\varepsilon) + 2\exp(-\varepsilon)$ and $I_2 = \exp(-2\varepsilon) + 2\exp(\varepsilon)$ where $\varepsilon$ is the Hencky strain. $\alpha$ is an adjustable parameter [Wagner et al. (1979)]. The material functions for shear with the double exponential damping function are provided by Kalyon et al. (1988) and were used in the analysis of the data of this study.

The uniaxial extensional stress growth behavior needs to be calculated via numerical integration,

$$[\tau_{11} - \tau_{22}](t) = \int_0^t \sum_i \frac{G_{oi}}{\lambda_i} \exp(-\frac{s}{\lambda_i}) h(s)(e^{2\varepsilon(s)} - e^{-\varepsilon(s)}) ds + \sum_i G_{oi} \exp(-\frac{t}{\lambda_i}) h(t)(e^{2\dot{\varepsilon}t} - e^{-\dot{\varepsilon}t}) \qquad (6)$$

where s is the elapsed time and where $G_{oi}$ is the relaxation strength, $G_{oi} = \eta_i/\lambda_i$. Another form of the nonlinear integral constitutive equation, which is very similar to the K-BKZ equation, is the Rivlin-Sawyers Equation, given here in the factorized form [Rivlin and Sawyers (1971)],

$$\tau(t) = \int_{-\infty}^{t} M(t-t')\left[\phi_1(I_1,I_2)\gamma_{[0]} + \phi_2(I_1,I_2)\gamma^{[0]}\right]dt' \qquad (7)$$

This equation is more general then the factored K-BKZ equation because, the kernel functions $\phi_1$ and $\phi_2$, are not derivatives of a strain-energy function with respect to the invariants, $I_1$ and $I_2$, and thus are easier to fit to experimental data [Macosko (1994)]. In fact, Equation 7 also reduces to the Wagner model with the double exponential damping function if one equates $\phi_2$ to zero and $\phi_1$ to f exp (-$n_1\gamma$) + (1-f) exp (-$n_2\gamma$). Papanastasiou et al. (1983) suggested equating $\phi_2$ to zero and $\phi_1$ to $1/(1+a(I_1-3)+b(I_2-3))$. Replacing M(t-t′) with the conventional linear viscoelastic fading memory function, Papanastasiou-Scriven-Macosko equation reduces to [Papanastasiou, (1984)],

$$\tau(t) = \int_{-\infty}^{t}\left[\sum_k G_i e^{-\frac{t-t'}{\lambda_k}}\right]\frac{a}{(a-3)+bI_1+(1-b)I_2}B_t(t')dt' \qquad (8)$$

For simple shear $I_1=I_2=3+\gamma^2$ the damping function reduces to the following sigmoidal function:

$$h(\gamma)=a/(a+\gamma^2) \qquad (9)$$

Sosky and Winter (1984) proposed a modification of the sigmoidal shear damping function, where the exponent of the shear strain becomes an additional material parameter, b, as:

$$h(\gamma) = \frac{1}{1+a\gamma^b} \qquad (10)$$

The functional dependence of h on $\gamma$ and the approximate time scale at which time and strain separability breaks down were investigated for monodispersed linear polymers by Doi and Edwards (1978) based on the reptation theory of deGennes (1971). It was determined that at times greater than the time, $t_{eq}$, it takes for a strand to retract back to its characteristic length of the equilibrium state, following a deformation of step strain nature, the separability of time and strain holds. On the other hand, at times smaller then $t_{eq}$, the strands keep on shrinking back within their tube-like region in which the strand is confined by entanglements with surrounding molecules. The damping function, which Doi and Edwards obtained, was approximated by Larson (1984) as,

$$h(\gamma) = \left(1 + \frac{1}{5}\gamma^2\right)^{-1} \tag{11}$$

**Estimation of the parameters**

**Relaxation Spectra**

The storage modulus, G', and loss modulus, G", data collected over a relatively broad range of frequencies, $\omega$, are generally employed to determine the relaxation spectra, $\eta_i$ (or $G_{oi}$) versus the relaxation time, $\lambda_i$ using linear and nonlinear regression analysis, while simultaneously adjusting both the relaxation times and strengths [Orbey and Dealy (1991) and Baumgaertel (1992)]:

$$G'(\omega) = \sum_i \frac{G_{oi}\lambda_i^2\omega^2}{1+\lambda_i^2\omega^2} \quad and \quad G''(\omega) = \sum_i \frac{G_{oi}\lambda_i\omega}{1+\lambda_i^2\omega^2} \tag{12}$$

Here, the characterization of the discrete relaxation spectra, was carried out in conjunction with Generalized Reduced Gradient type of nonlinear optimization by using a pattern search method, which minimizes a least squares objective function, F,:

$$F = \sum_{k=1}^{N}[((G'_{k,\exp} - G'_{k,fit})/G'_{k,\exp})^2 + ((G''_{k,\exp} - G''_{k,fit})/G''_{k,\exp})^2] \tag{13}$$

where N is the number of data points, $(G'_{k,\exp}, G''_{k,\exp})$ available from the small-amplitude oscillatory shear experiments. $G'_{k,fit}$ and $G''_{k,fit}$ denote the best-fit values on the basis of Equation (13). Both the relaxation times, $\lambda_i$, and relaxation strengths, $G_{oi}$, were allowed to float during the error minimization. The conjugate search algorithm was utilized to determine the direction to search for the error minimization. The derivatives of the objective functions were obtained by the forward difference method.

**Characterization of the parameters of the damping function:**
**Step strain flow for the determination of the relaxation modulus as a function of time and strain:**

Stress relaxation following a sudden shearing displacement (also referred to as the step-strain stress relaxation) is generally used to characterize the damping function [Einaga et al. (1971); Lodge (1975); Larson (1985); Kalyon et al. (1988); Kasehagen and Macosko (1998); Partal et al. (1999); Madiedo et al. (2000); Nishioka et al. (2000); Rubio and Wagner (2000); Gevgilili and Kalyon (2001); Carriere et al. (2002)]. In this experiment the specimen is held between a cone and a disk or in between two parallel disks under isothermal conditions. A shear strain $\gamma_0$ is introduced by applying a large and presumably constant shear rate at time $t'=t_0$ for a short duration of time. The idealized strain history is:

$$\gamma_{xy}(t')=0 \quad \text{for} \quad t' \geq t_0 \quad (14)$$
$$\gamma_{xy}(t')=\gamma \quad \text{for} \quad t' < t_0$$

Upon this sudden imposition of the shear strain the measurement of the decay of the shear stress, $\tau_{xy}(t)$, with time provides the relaxation modulus $G(t, \gamma)$:

$$G(t, \gamma) = \tau_{xy}(t)/\gamma \quad (15)$$

For relatively small shear strains the relaxation modulus is found to be independent of the imposed shear strain i.e.,

$$\lim_{\gamma \to 0} G(t, \gamma) = G(t) \tag{16}$$

In the limit of the linear viscoelasticity the relaxation modulus can be predicted using the relaxation spectra determined from the small-amplitude oscillatory shear experiments. For example, for the generalized Maxwell model the shear relaxation modulus in the linear region becomes [Bird et al. (1987)]:

$$G(s) = \sum_{i=1} G_{oi} e^{(-s/\lambda_i)} \tag{17}$$

where $s = (t - t')$ is the elapsed time.

The step strain experiment provides a convenient way to characterize the damping function, $h(\gamma)$, that can be obtained at each value of the strain, $\gamma$, by the vertical shift of the $G(t, \gamma)$ onto the linear viscoelastic modulus $G(t)$:

$$h(\gamma) = G(t, \gamma)/G(t) \tag{18}$$

However, the step strain experiment, which by definition involves the introduction of the strain, in the fastest manner possible, i.e., in milliseconds, can lead to the slip of the melt at the wall of the rheometer. This was indeed shown to be the case for the high density polyethylene of this study, which exhibited strong wall slip during the step strain experiment over a broad range of imposed strains [Gevgilili and Kalyon (2001)]. Thus other ways of the characterization of the damping function using other material functions, had to be sought. Examples of such methods available in the literature include the determination of the damping function parameters from the combination of shear stress growth and steady shear viscosity data [Kasehagen and Macosko (1998)] and combination of shear viscosity and first normal stress coefficient data [Maia (1999)]. On the other hand, the second material of this study, the thermoplastic elastomer did not exhibit strong wall slip during the step strain experiment allowing its damping function to be characterized using the step strain experiment along with the other inverse problem solution based methods also applied to HDPE.

**Experimental:**

**Materials:**

The experiments were carried out using two resins, a high-density polyethylene, HDPE and a thermoplastic elastomer, TPE. HDPE is manufactured by Exxon-Mobil Chemical Company (HDA 601). It has a melt flow index of 0.55 g/10 min (ASTM D-1238) and a density of 0.933 g/cm$^3$ (ASTM D-1505). The TPE is an oxetane-based alternating block copolymer, available from ATK Thiokol Propulsion Corporation of Promontory, Utah. The hard blocks of this TPE consist of [3, 3-bis (azidomethyl) oxetane, BAMO] and the soft blocks are (3-azidometyl-3-metyloxetane, AMMO). This TPE was included in this study upon our observation that the melt of this polymer exhibited unusually high stickiness to metal surfaces that became apparent during the loading of rheometers and processing equipment. Subsequent studies indicated that the TPE did not exhibit strong wall slip (slip which can be detected easily using the straight-line marker technique [Kalyon et al. (2003)]) over a broad range of strain rates and strains in simple shear flow. The TPE also did not exhibit flow instabilities in capillary rheometry. Additional information on these two polymers can be obtained elsewhere [Gevgilili and Kalyon (2001) and Kalyon et al. (2003)].

**Equipment and procedures**

For the step strain, oscillatory shear, stress growth and relaxation experiments the Advanced Rheometric Expansion System (ARES) rheometer available from TA Instruments was used. Various features of this rheometer, as well as possible error sources, are discussed in Gevgilili and Kalyon (2001). For cone-and-plate and parallel disk experiments the diameters of the cones and disks were 8 and 25 mm and the cone angle was around 0.1 radians (the cone angle of each cone used was measured precisely). The rheometer was modified by installing an imaging window into the wall of the environmental chamber. This facilitated the continuous monitoring of the deformation at the free surface of the melt [Gevgilili and Kalyon (2001)].

The step strain data need to be corrected for the non - homogeneous strain when parallel disk fixtures are employed. The following equation [Soskey et. al. (1984)] is used to correct the apparent relaxation modulus $G_a(t, \gamma_R)$:

$$G(t, \gamma_R) = G(t, \gamma_R) = \left(1 + \frac{1}{4} \frac{\partial \ln G_a(t, \gamma_R)}{\partial \ln \gamma_R}\right) \quad (19)$$

where $G(t, \gamma_R)$ is the corrected modulus at the strain corresponding to the edge of the disc, $\gamma_R$, and $G_a(t, \gamma_R)$ is given by:

$$G_a(t, \gamma_R) = 2T(t, \gamma_R) / \pi R^3 \gamma_R \quad (20)$$

where, $T(t, \gamma_R)$ denotes the torque exerted on the upper disk.

The shear viscosity material functions of the two materials in the relatively high apparent shear rates of 10 to 50 s$^{-1}$ for the high density polyethylene, HDPE, and 10-200 s$^{-1}$ for the thermoplastic elastomer were also characterized by using an Instron Capillary Rheometer. The uniaxial extensional stress growth behavior of the two materials were characterized using a Meissner-type uniaxial extensional rheometer [Meissner (1994)]. The extensional rheometer was equipped with a CCD camera and appropriate image analysis capabilities to allow the determination of the true strain rate from the dimensions of the specimen as a function of time during the uniaxial stretch [Wagner et al., 2002)].

**Results and Discussion:**

**Determination and testing of parameters for TPE**

The TPE was subjected to small-amplitude oscillatory and steady shear experiments at 80, 100 and 120°C. Assuming that the TPE is a thermo-rheologically simple fluid the variation of the temperature corresponds to a shift in the time scale, represented by the shift factor, $a_T$, given as:

$$a_T(T) = \exp\left[\frac{E_a}{R}\left(\frac{1}{T} - \frac{1}{T_{ref}}\right)\right] \quad (21)$$

where $E_a$ in the activation energy, R is the Gas constant, T is temperature and $T_{ref}$ is the reference temperature. The dynamic and loss modulus data were fitted to generate the discrete relaxation spectra given in Figure 1. The relaxation times are in the $5.7*10^{-5}$ to 837 s range and the corresponding relaxation modulus values are in the $1.85*10^6$ Pa to 46.9 Pa range. The time-temperature superposition is observed to hold well and the activation energy is 17.9 kcal/mol (Figure 1).

Figure 2 shows the relaxation modulus versus time and strain behavior of the TPE in the strain range of 20 to 900%. The time-strain separability appears to hold. No evidence of strong wall slip was observed for the TPE during the step strain experiment [Kalyon and Gevgilili (2003)]. The absence of strong wall slip in step strain allowed us to include the step strain data as part of the procedure for the characterization of the damping function parameters for the TPE. The adequate fit of Equation 17 to the relaxation modulus data in the linear viscoelastic range is also shown in Figure 2.

The best-fits of various types of damping functions (Equations 3, 9-11) to the experimental step strain data of the TPE are shown in Figure 3. Various forms of damping function, i.e., the double-exponential damping function (Equation 3), sigmoidal Papanastasiou-Scriven-Macosko damping function (Equation 9), with one adjustable parameter, i.e., (a), the Sosky-Winter sigmoidal equation, (Equation 10), which has two adjustable parameters, (α,β), and the simplified Doi and Edwards damping function (Equation 11), were fitted to the relaxation modulus versus strain data of TPE. Overall, the damping function values cannot be adequately fitted with a single parameter and Equations 3 and 10, which contain multiple parameters, fit the damping function better.

The comparisons of the predictions versus the shear viscosity data of TPE are shown in Figure 4 for the four different types of damping function. The observed poor representations of the damping function, h, at high strain levels for the Papanastasiou-Scriven-Macosko and Doi-Edwards equations, give rise to predictions of shear viscosity values, which are significantly smaller than the experimentally-determined values. The predictions associated with damping functions, which involve more than or equal to two parameters, agree with the experimental data. Similar conclusions could be drawn from the comparisons of the experimental and predicted values of the other material functions including the first normal stress difference and shear stress growth and relaxation upon cessation of steady shear data collected under steady shearing at various shear rates.

For example, shear stress growth data of TPE upon the imposition of the shear rates of 0.05 and 1 $s^{-1}$ are shown in Figure 5. In these experiments the TPE stays intact within the gap and does not exhibit wall slip for strains as high as 100. Upon larger strains, minor sample ejection out of the gap and edge distortions are observed to occur with concomitant decreases of the shear stress values. At these shear rates the agreement between the predictions of the double exponential damping function and the experimental data is excellent. The sigmoidal damping function with two parameters (Equation 10) underpredicts the shear stress values after the 10 s of deformation. Another interesting feature of this plot is the prediction of a slight shear stress overshoot by the double exponential, Papanastasiou-Scriven-Macosko and Doi-Edwards damping functions and the lack of it in the predictions of the Sosky-Winter damping function. We will revisit the stress overshoot issue later.

The uniaxial extensional stress growth experimental data and the predictions of the integral-type constitutive equation in conjunction with h=1 are shown in Figure 6. The true Hencky strain rates were determined upon the analysis of the time-dependent widths of the strands with time during uniaxial stretch. The uniaxial extensional flow data of TPE exhibits even a greater degree of strain hardening than what Wagner model with h=1 predicts. The role of the damping function, h, is to reduce the strain hardening behavior and the inclusion of any type of damping function to lead to h<1 will increase the differences between the predictions of the single-integral constitutive equation and the experimental uniaxial extensional growth data.

Overall, these results suggest that for the TPE, which does not exhibit strong wall slip over a broad range of strains and shear rates, the appropriate selection of the form of the damping function and the step strain experiment are adequate to predict various material functions in shear successfully, as consistent with the behavior observed for other resins [Kalyon et. al. (1988)]. However, problems persist in the prediction of the extensional flow behavior regardless of which form of the damping function is used. In the next section the resources that are necessary for conditions at which the damping function parameters cannot be determined from the step strain experiment due to strong wall slip are discussed.

**HDPE (HDA-601)**

The small-amplitude oscillatory shear and steady shear flow behavior of the HDPE was characterized at 170, 180°C and 190°C. The uniaxial extensional stress growth behavior of the HDPE was characterized at 190°C. The dynamic properties, i.e., storage, G', and loss modulus, G", values could be fitted adequately using a discrete relaxation spectrum with eight modes, the relaxation times ranging from $5.8*10^{-5}$ to 35 seconds with a corresponding relaxation strength range of $7.8*10^6$ to 160 Pa (Figure 7). The time-temperature superposition furnished an activation energy value of 6.07 kcal/mol in the 170-190°C range at a reference temperature of 190 ºC.

Step strain experiments were earlier performed for this HDPE up to a strain of 400%. Strong wall slip was observed for strains that were as low as 100% [Gevgilili and Kalyon (2001)]. Upon the corrections for wall slip the true strain values imposed during step strain were found to be significantly smaller than the apparent strain values and thus the resulting relaxation data were reduced to a small range of strains, as shown in Figure 8. Upon correcting the step strain data for wall slip, the maximum true shear strain achieved was only about 140% (in comparison to the apparent strain value of 400%) and the bulk of the corrected data were in the linear viscoelastic range. The best-fits for the damping function using the step strain data thus become dependent on only the few data points available in the nonlinear region and thus all of the fits become unacceptable.

If the step strain data become unreliable due to wall slip what recourse is available to fit the parameters of the damping function? Can one utilize other material functions characterized over ranges of shear rate in which wall slip effects are negligible?

**Utilization of other Material Functions to determine the damping function instead of the step strain flow data that are unreliable due to wall slip; Solution of the inverse problem:**

The characterization of the damping function upon the solution of the inverse problem using data from material functions such as shear stress and first normal stress difference growth and relaxation, collected upon the initiation and cessation of steady shear, present several advantages: (1) the experiments can be performed at sufficiently low deformation rates, which would minimize or eliminate the contribution of strong wall slip, (2) only a single experiment with reliable data may be sufficient to fit the parameters of the damping function compared to the multiple experiments which are necessary in the case of step strain experiments. (3) Furthermore, the experiments necessary to generate time-dependent shear stress and first normal stress difference data are simpler in comparison to those necessary to generate the step strain experiment, which need to deform the sample at very high strain rates and over very brief durations of time, with the associated difficulties in generating the necessary acceleration and the deceleration of the tools [Gevgilili and Kalyon (2001)].

In the procedures which were set up for the determination of the damping function parameters starting from the data associated with one material function or various combinations, the damping function parameters were allowed to float while the error of a least squares objective function was minimized with respect to the available experimental data. This is basically the solution of the inverse problem. Only the double exponential damping function (Equation 3) was used for this purpose due to the relatively simpler mathematical structure of the inverse problem for the double exponential type damping function. The search routine is executed to find the parameters of the double exponential damping function, i.e., f, $n_1$ and $n_2$, which minimize the objective function, F. The partial derivatives of the objective function were evaluated by the forward difference method,

$$\frac{\partial F(f,n_1,n_2)}{\partial f} = \frac{F_{f+\Delta f,n_1,n_2} - F_{f,n_1,n_2}}{\Delta f}$$

$$\frac{\partial F(f,n_1,n_2)}{\partial n_1} = \frac{F_{f,n_1+\Delta n_1,n_2} - F_{f,n_1,n_2}}{\Delta n_1} \quad (22)$$

$$\frac{\partial F(f,n_1,n_2)}{\partial n_2} = \frac{F_{f,n_1,n_2+\Delta n_2} - F_{f,n_1,n_2}}{\Delta n_2}$$

The calculations were made for both the TPE and the HDPE. The values of the objective functions, i.e., the normalized sum of the squares of the differences between the predicted and experimental values:

$$F = \sum_{k=1}^{N}((D_{k,\exp} - D_{k,fit})/D_{k,\exp})^2 \quad (23)$$

for the TPE and the HDPE are reported in Tables 1 and 2, respectively, based on the use of different material functions for the fitting of the parameters of the damping function. For TPE, the parameters of the damping function determined on the basis of the fitting of various material functions (and their combinations) compared very favorably with those determined from the relaxation modulus data obtained from the step strain experiment as shown in Table 1. Thus, for the TPE the step strain experiment was free of the wall slip effects and generated the double exponential damping function parameters (Equation 3) as f=0.51, $n_1$=0.01, $n_2$=0.49. As shown in Table 1, the solutions of the inverse problem for the parameters of the damping function obtained from the inverse solution of other material functions generated damping function parameters, which were very close to those determined with the step strain experiment. For example, the inverse problem solution using only the shear stress growth data collected at shear rates of 0.05 s$^{-1}$ and 1 s$^{-1}$ generated the damping function parameters of f=0.6, n$_1$=0.01, n$_2$=0.6, which are indeed very close to those determined using the step strain experiment. The parameters determined from the other tests are also in the same range.

The values of the objective function reported in Table 1 suggest that the parameters determined from material functions such as the shear stress and first normal stress difference relaxation and the growth are indeed suitable to predict the other material functions of the TPE. For example, the typical shear stress and first normal stress difference growth predictions of the Wagner constitutive equation, using the parameters of the double exponential damping function determined upon the inverse problem solution using only the first normal stress difference relaxation data of the TPE, are shown in Figures 9 and 10. The good agreements between the predicted values of the material functions and the experimental data suggest again that such procedures can serve as a substitute for the determination of the parameters of the damping function from the step strain experiment.

The typical best-fit of the first normal stress difference relaxation data of the TPE, collected upon the cessation of steady shear at 1 s$^{-1}$, is shown in Figure 11. The set of damping function parameters which are determined on the basis of the fit of this set of data are f=0.49, n$_1$=0.11 and n$_2$=1. The best fit of the damping function is seen to be acceptable. There is adequate agreement between the predictions using this set of damping function parameters versus the experimental data of the shear viscosity, first normal stress difference and shear stress relaxation following the cessation of steady shear at 1 s$^{-1}$ data as shown in Figures 12-14.

Now let us apply the inverse problem solution procedure to the determination of the parameters of the double exponential damping function (Equation 3) for the high density polyethylene melt for which reliable step strain data could not be obtained due to the wall slip. The damping function parameters obtained upon the inverse problem solution using the shear stress growth and shear stress relaxation data fall in the same range and approach the behavior of a damping function with a single exponential (Table 2). The ability to predict the growth and steady values of the first normal stress difference values for HDPE is limited, whereas the other material functions, i.e., the shear stress growth and relaxation, steady shear viscosity and first normal stress difference relaxation could be accurately predicted, as suggested by the relatively low values of the objective function (Equation 23).

For the HDPE the agreements between the predictions and the experimental values of the first normal stress difference growth are not as good as those observed for the TPE possibly indicating the complex role wall slip plays in the characterization of the first normal stress difference. Overall, the suggested inverse problem solution based method of the estimation of the parameters of the damping function from one or more experiments like shear stress or the first normal stress difference relaxation upon the cessation of the steady shear significantly simplifies the characterization of the parameters of the damping function.

The uniaxial extensional flow data of HDPE are shown in Figure 15 at two Hencky strain rates, 0.055 and $0.6s^{-1}$. The experimental data did not exhibit strain hardening within the time scale of examination. For the case of the HDPE the use of h=1 predicts significant strain hardening which needs to be dampened by the inclusion of the damping function, in a form suitable for extensional flow. The inclusion of the damping function of the Wagner model using the generic parameters for polyethylene as suggested by Wagner (1977) appears to help up to a certain time value, however, the predictions are still inadequate to represent the uniaxial extensional data.

An important issue associated with the inverse problem methodologies suggested here is the overall lack of uniqueness. Different sets of parameters can represent the data equally well and there is no recourse available to generate a unique set of parameters. This is important because subtle changes in the parameters used (which are well within the confidence intervals of any experimental procedures utilized) can give rise to misleading results. Let us consider, for example, whether the shear stress or first normal stress difference overshoots are predicted or not. Kalyon and Gevgilili (2003), have shown experimentally, for three polymers that they have investigated in detail, that the shear stress and first normal stress difference overshoots could be artifacts of wall slip. For example, the TPE of this study, which does not exhibit strong wall slip over a broad range of shear rates and strains, also does not exhibit overshoots. On the other hand, the apparent overshoots in shear stress and the first normal stress difference which were observed for the HDPE of this study were determined to be direct consequences of wall slip [Kalyon and Gevgilili (2003)]. Does the Wagner postulate of the K-BKZ equation considered here consistently predict overshoots?

Earlier it was shown that different types of damping functions give rise to different behavior in stress overshoots of the TPE (see Figure 5). An additional parametric study using different sets of the parameters of the relaxation spectra and damping function indicated that even small changes in the parameters were sufficient to alter the qualitative nature of the predictions from the absence to the presence of stress overshoots. This typical behavior is shown in Figure 16, where predictions of the shear stress growth, using two different sets of relaxation spectra, which equally fit the dynamic data are provided. One of the two sets predicts a shear stress overshoot, whereas the second set does not. Similar observations were made on the basis of the different sets of parameters used to represent the damping function.

**Conclusions:**

The conventional approach for the determination of the parameters of the damping function for strain-time separable constitutive equations involves the use of the step strain experiment in which a very high rate of deformation is applied and then ceased within milliseconds. Such high rates of deformation can give rise to wall slip for some melts which can preclude the use of the step strain data. Here the solution of the inverse problem, using various material functions in shear, is suggested for the characterization of the parameters of the damping function when the step strain data are not useable. However, the lack of uniqueness of the parameters, regardless of how the relaxation spectra and the damping function parameters are characterized, is an outstanding serious issue and can lead to unrealistic predictions.

**List of Tables:**

**List of Figures:**

14. The experimental data and predictions (damping function obtained from first normal stress difference relaxation data) of the shear stress relaxation of HDPE at 190°C.
15. The experimental data and predictions of the uniaxial extensional viscosity growth of HDPE at 190°C
16. Predictions of the shear stress growth of the TPE at 1 s-1 using two different sets of relaxation spectra and similar damping function parameters of f=0.51, $n_1$=0.05, $n_2$=0.49.

**Table-1: Objective function values using different material functions for TPE.**

| Error minimized for | f | n1 | n2 | Shear Stress Growth | First Normal Stress Difference Growth | Shear Stress Relaxation | First Normal Stress Difference Relaxation | Steady Viscosity | Steady First Normal Stress Difference | Total Error |
|---|---|---|---|---|---|---|---|---|---|---|
| Step Strain | 0.51 | 0.01 | 0.49 | 1.05 | 1.09 | 0.47 | 0.76 | 0.18 | 6.23 | 9.8 |
| All Data | 0.77 | 0.01 | 0.90 | 1.00 | 0.68 | 0.34 | 0.39 | 0.19 | 0.57 | 3.2 |
| Shear Stress Growth | 0.60 | 0.01 | 0.47 | 0.95 | 0.68 | 0.84 | 2.17 | 0.42 | 9.14 | 14.2 |
| First Normal Stress Difference Growth | 0.73 | 0.01 | 0.66 | 1.02 | 0.54 | 0.82 | 0.94 | 0.40 | 2.94 | 6.7 |
| Shear Stress Relaxation | 0.60 | 0.02 | 0.02 | 1.46 | 1.32 | 0.20 | 1.51 | 0.30 | 0.40 | 5.2 |
| First Normal Stress Difference Relaxation | 0.67 | 0.01 | 0.01 | 1.68 | 1.19 | 0.74 | 0.25 | 0.65 | 0.72 | 5.2 |

Errors Generated on individual Material Functions

**Table-2: Objective function values using different material functions for HDPE.**

| Error minimized for | f | n1 | n2 | Shear Stress Growth | First Normal Stress Difference Growth | Shear Stress Relaxation | First Normal Stress Difference Relaxation | Steady Viscosity | Steady First Normal Stress Difference | Total Error |
|---|---|---|---|---|---|---|---|---|---|---|
| All Data | 0.11 | 0.05 | 1.00 | 0.39 | 33.90 | 2.25 | 1.31 | 0.35 | 0.50 | 38.2 |
| Shear Stress Growth | 0.70 | 0.23 | 0.23 | 0.13 | 60.71 | 0.74 | 1.50 | 0.28 | 0.08 | 63.4 |
| First Normal Stress Difference Growth | 0.91 | 1.00 | 1.00 | 0.52 | 32.03 | 6.27 | 9.73 | 0.94 | 1.99 | 49.5 |
| Shear Stress Relaxation | 0.98 | 0.25 | 0.00 | 0.13 | 59.64 | 0.10 | 31.44 | 1.02 | 29.70 | 92.3 |
| First Normal Stress Difference Relaxation | 0.49 | 0.11 | 1.00 | 0.17 | 46.11 | 0.41 | 0.08 | 0.27 | 0.04 | 47.0 |
| All-Normal Stress Difference Growth | 0.32 | 0.09 | 0.44 | 0.1 | 55.4 | 0.3 | 0.1 | 0.3 | 0.04 | 56.2 |

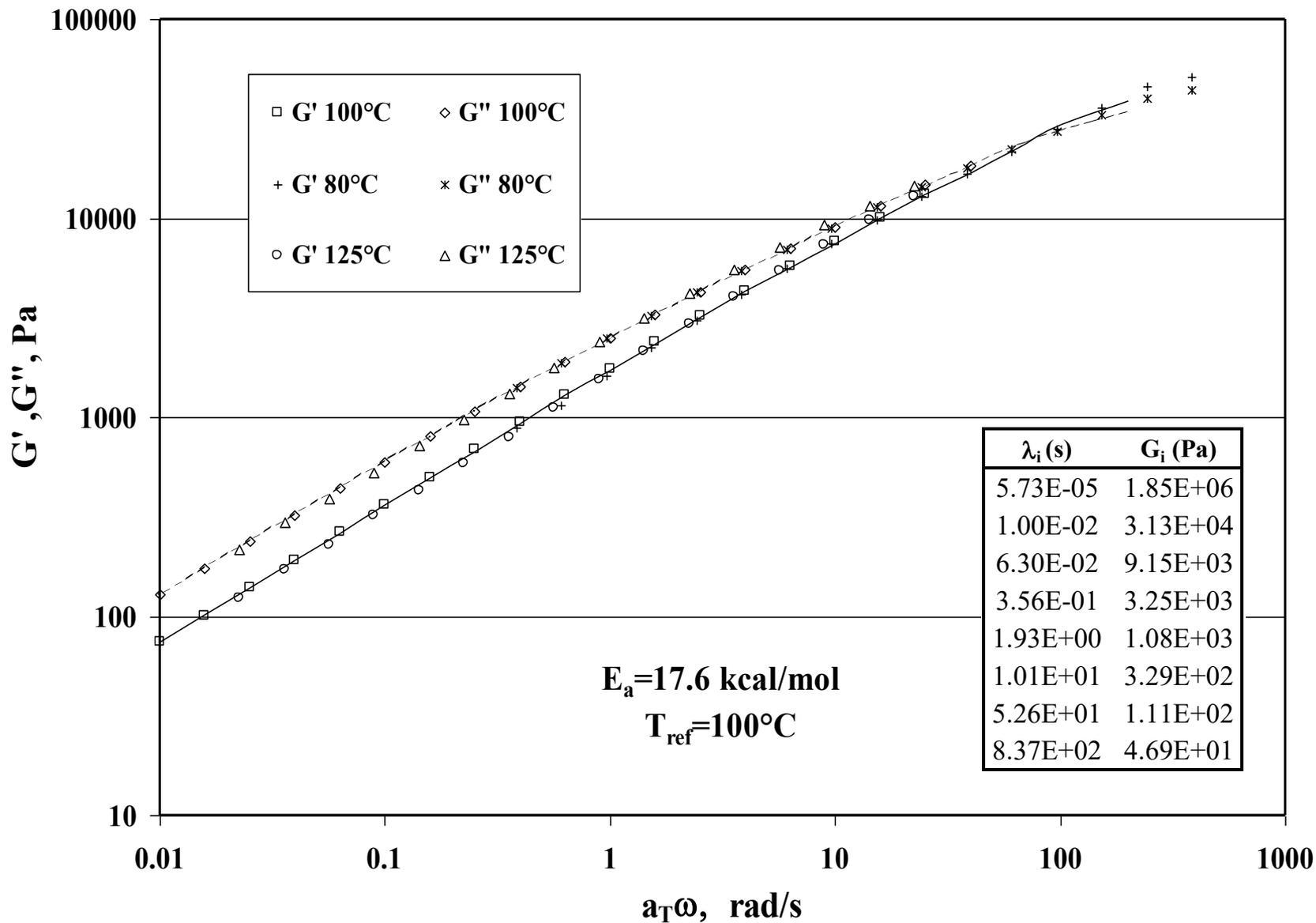

**Figure 1: The loss and storage modulus of TPE at various temperatures shifted to the reference temperature of 100°C.**

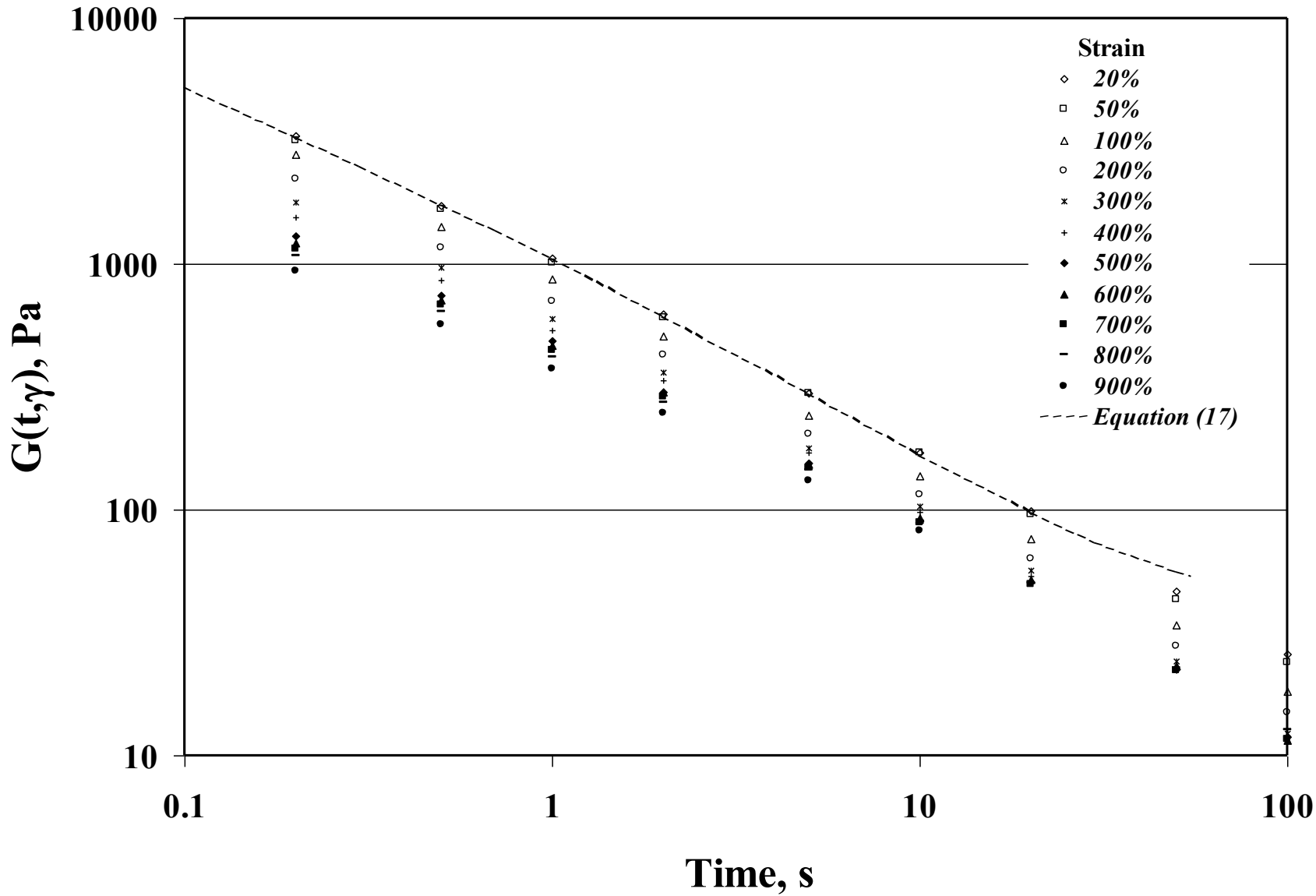

**Figure 2: Relaxation modulus of TPE as a function of strain and time**

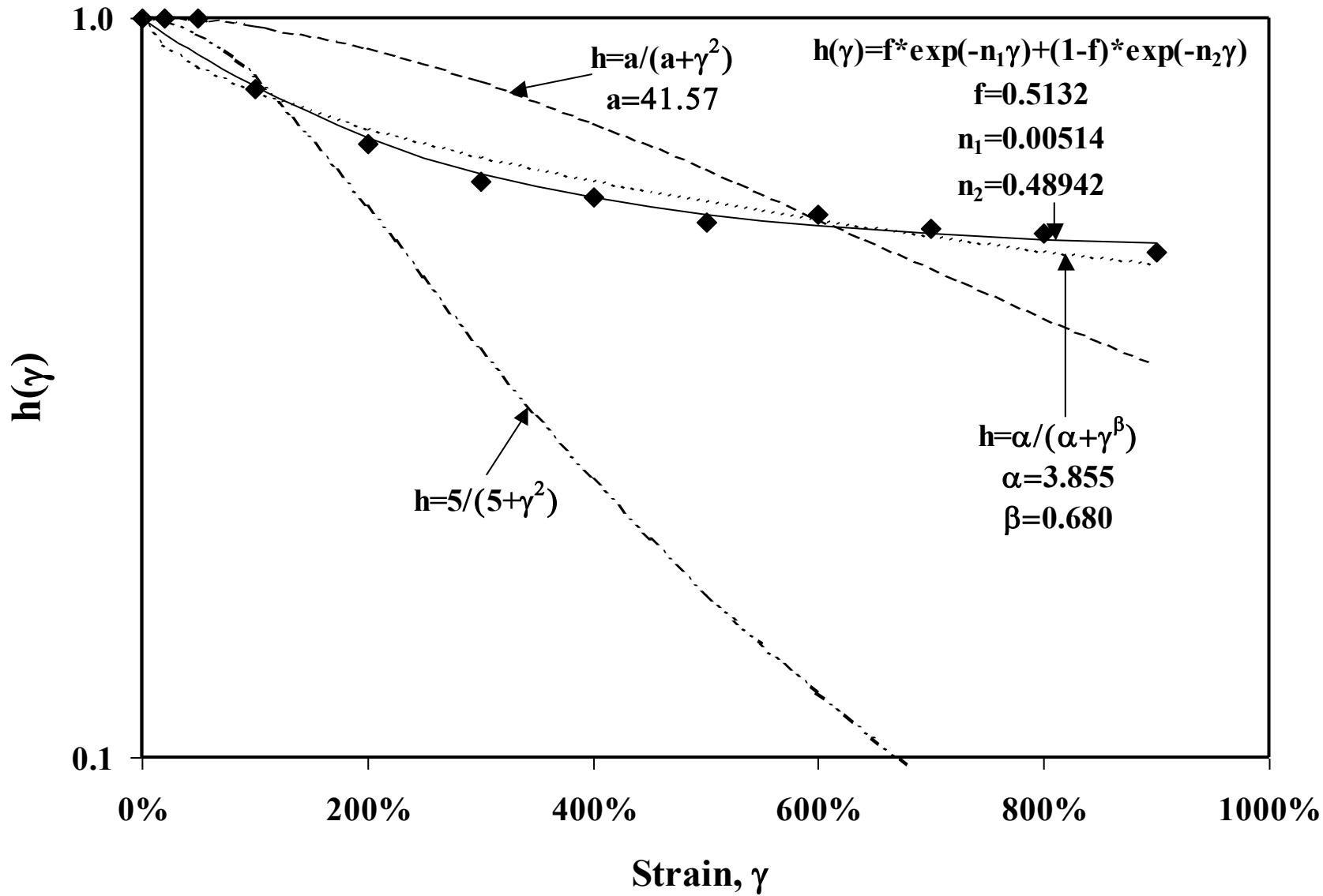

Figure 3: Various forms of damping functions are used to represent the step strain data of the TPE at 100°C

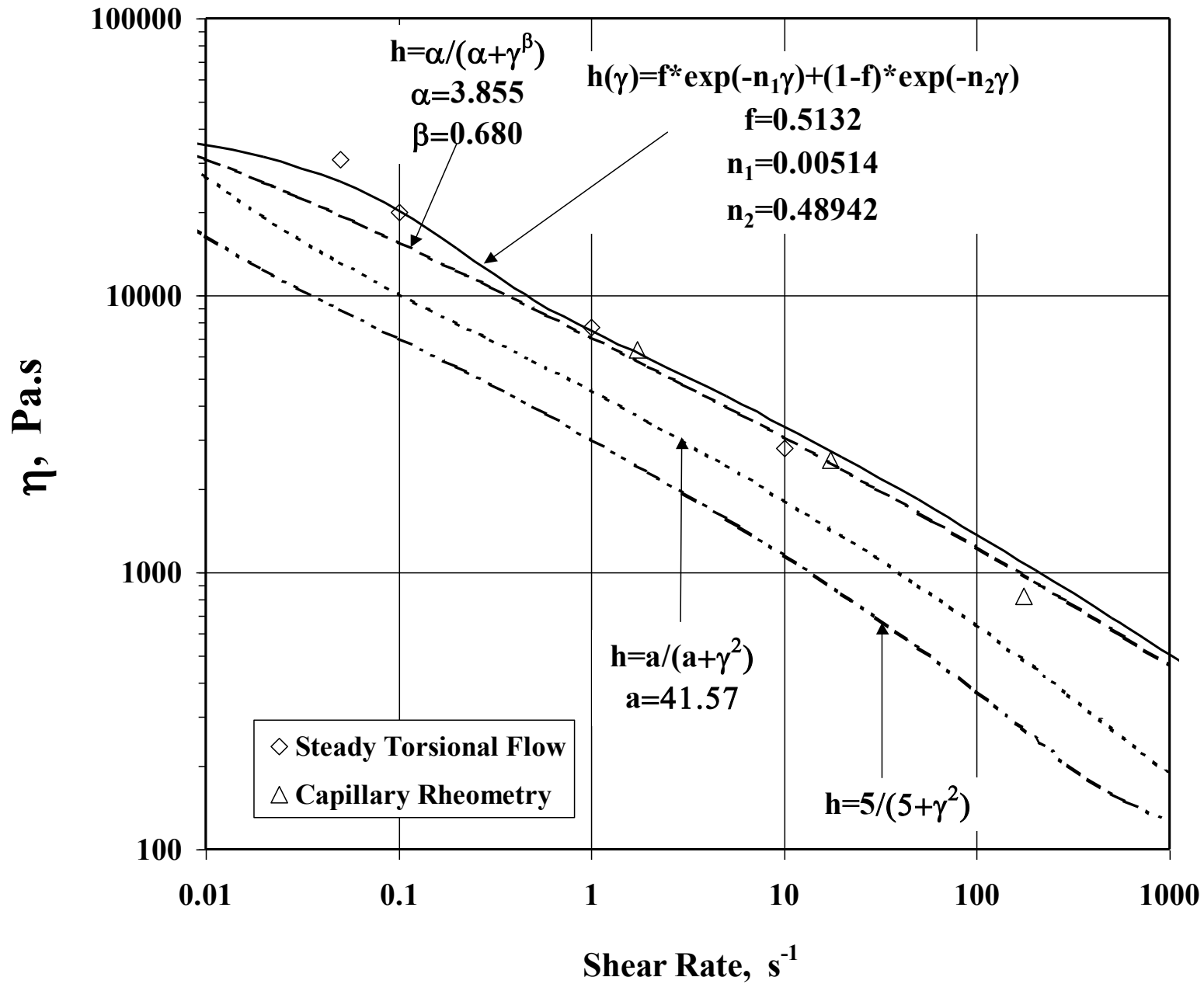

**Figure 4: The experimental data and predictions of the shear viscosity of the TPE at 100°C**

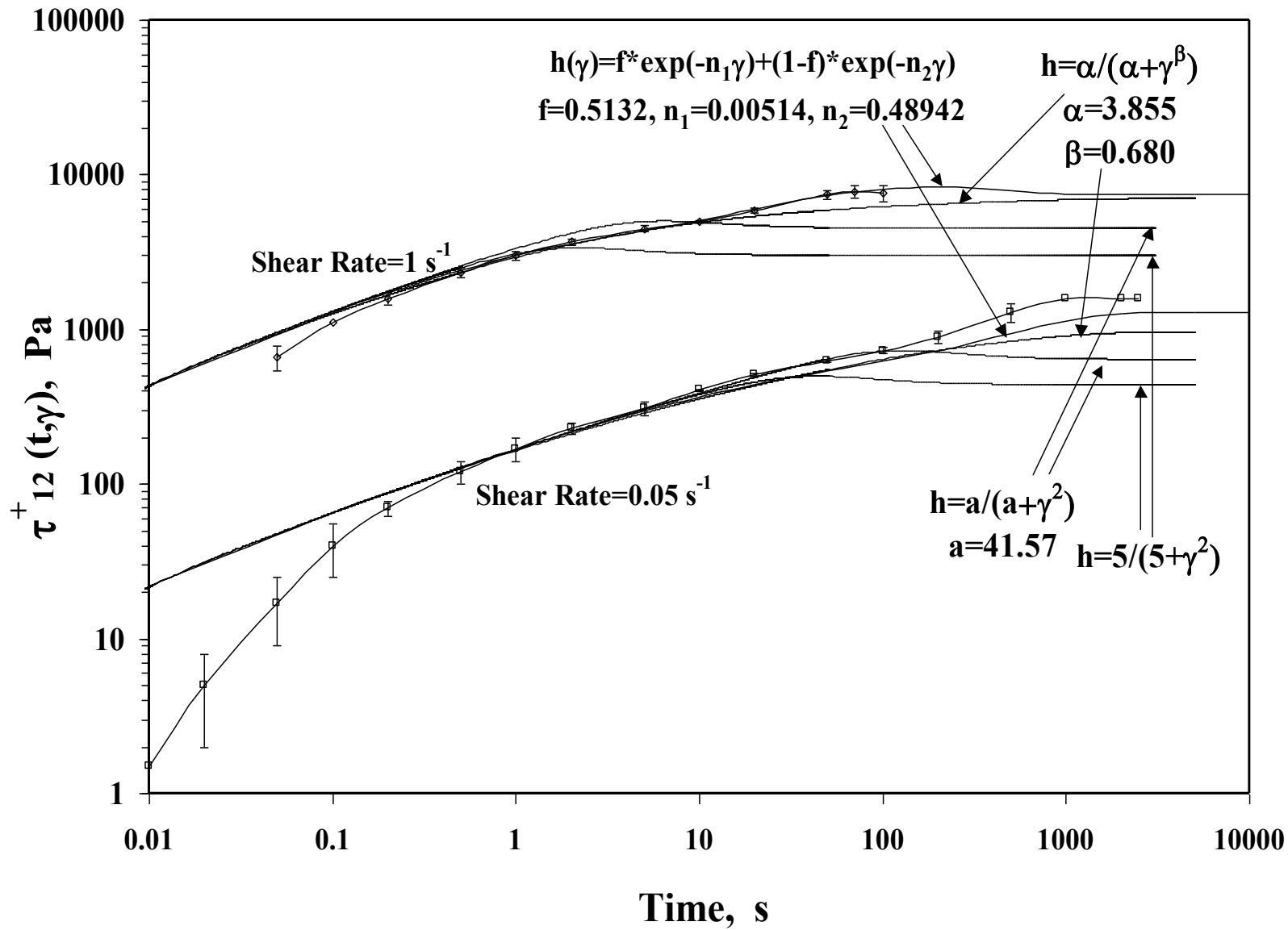

**Figure 5: The experimental data and predictions of the shear stress growth behavior of the TPE at 100°C.**

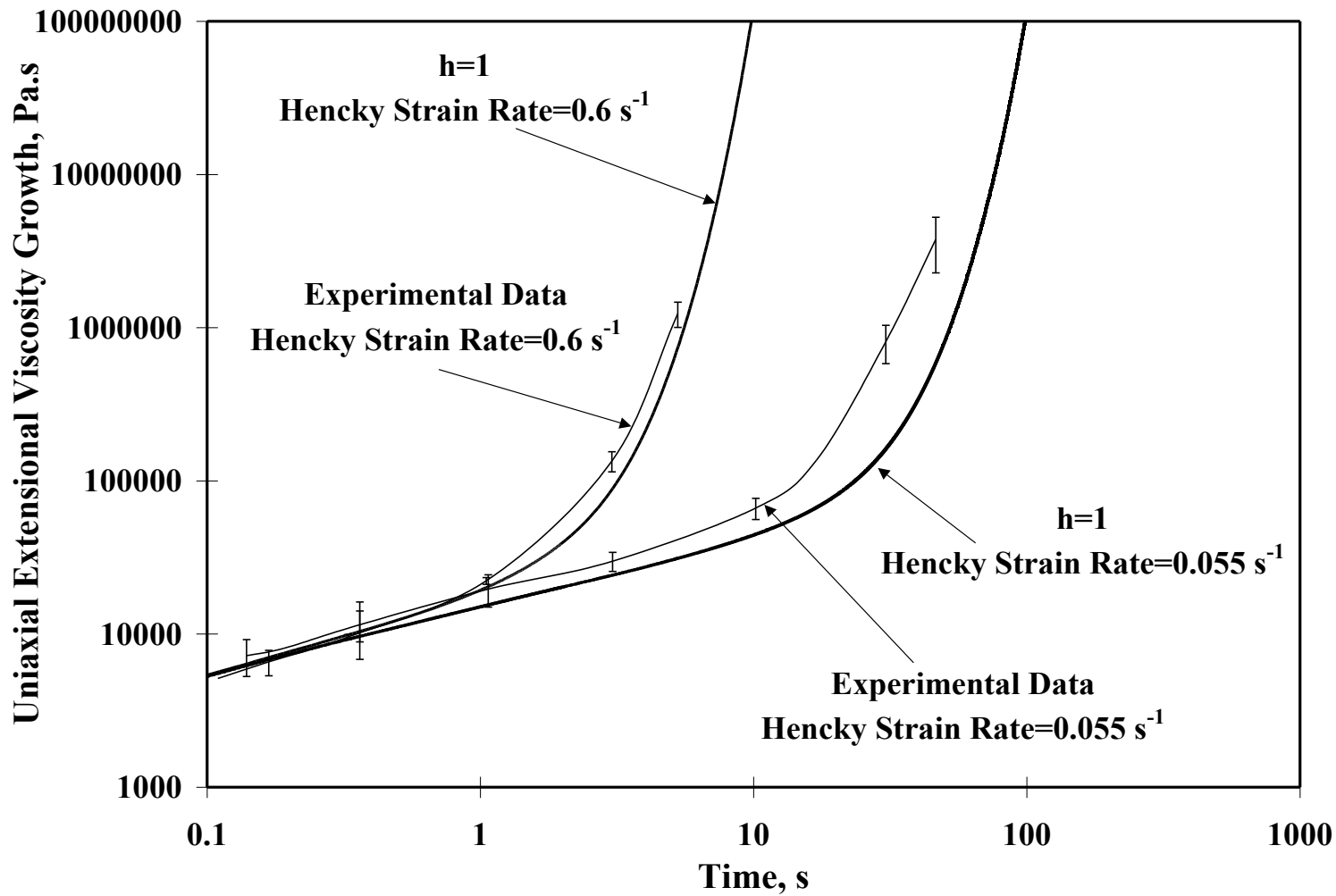

**Figure 6: The experimental data and predictions of the uniaxial extensional viscosity growth of the TPE at 100°C.**

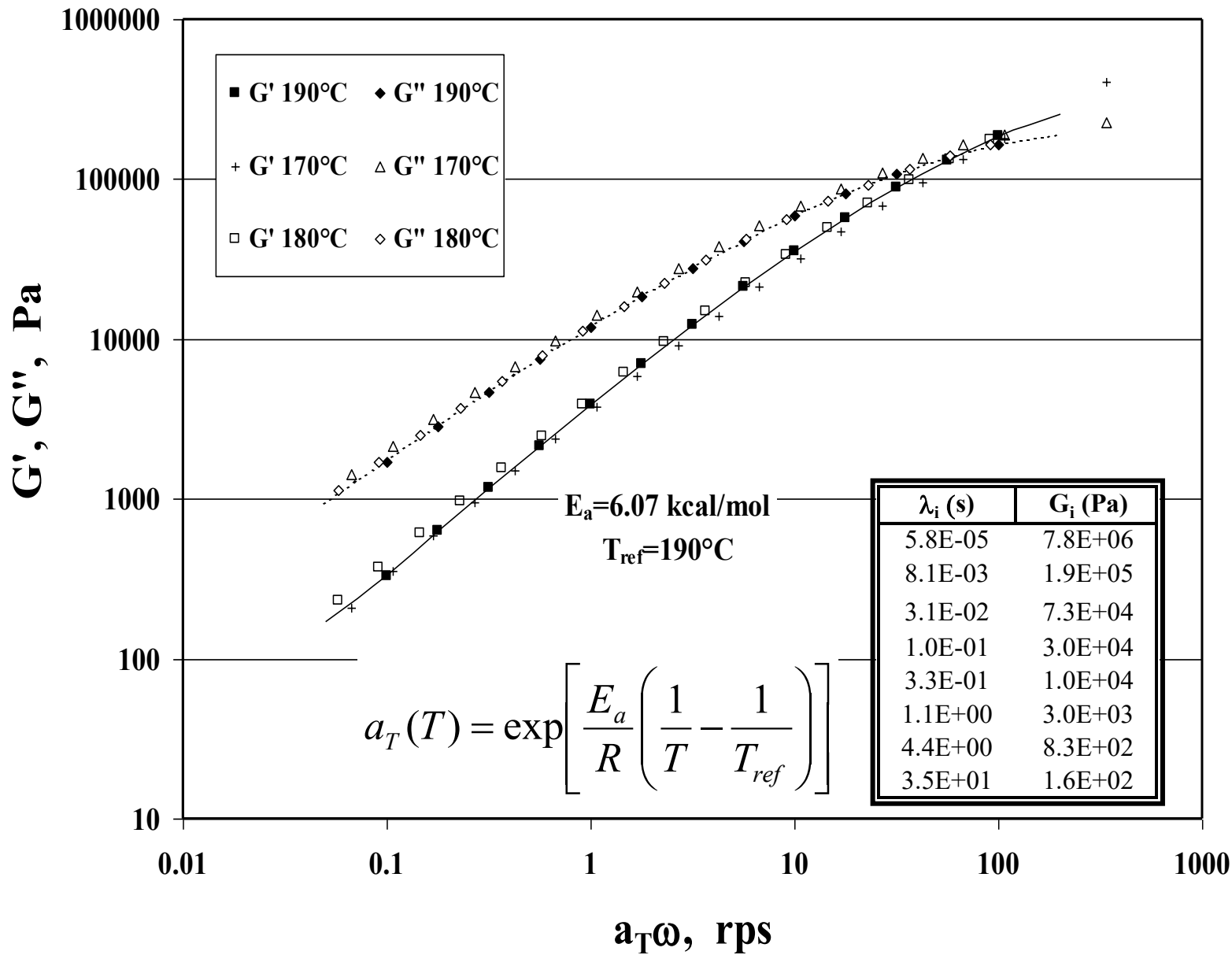

**Figure 7: The loss and storage modulus of HDPE at various temperatures shifted to the reference temperature of 190°C.**

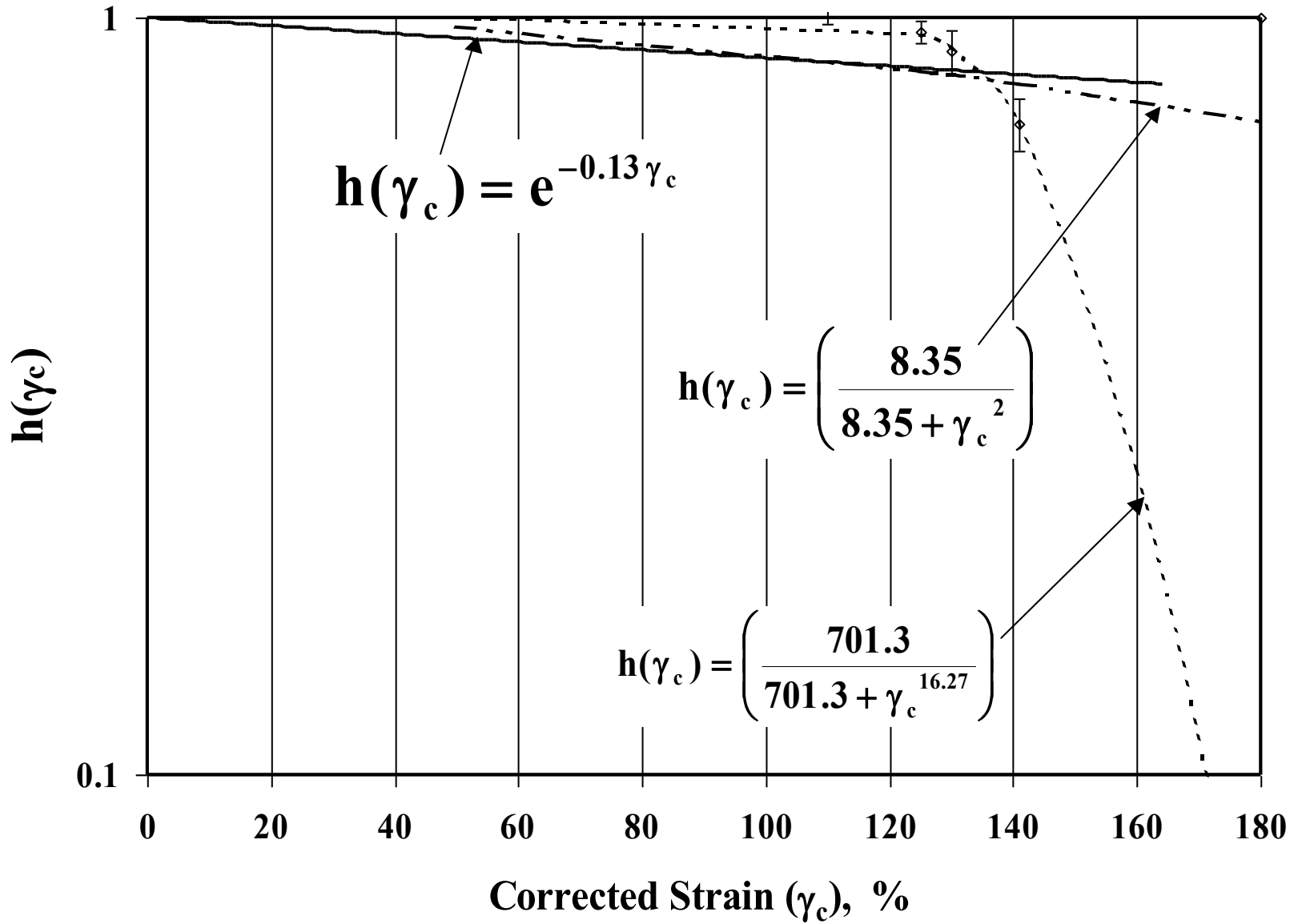

**Figure 8: Various forms of damping functions are used to represent the slip corrected step strain data of HDPE at 190°C.**

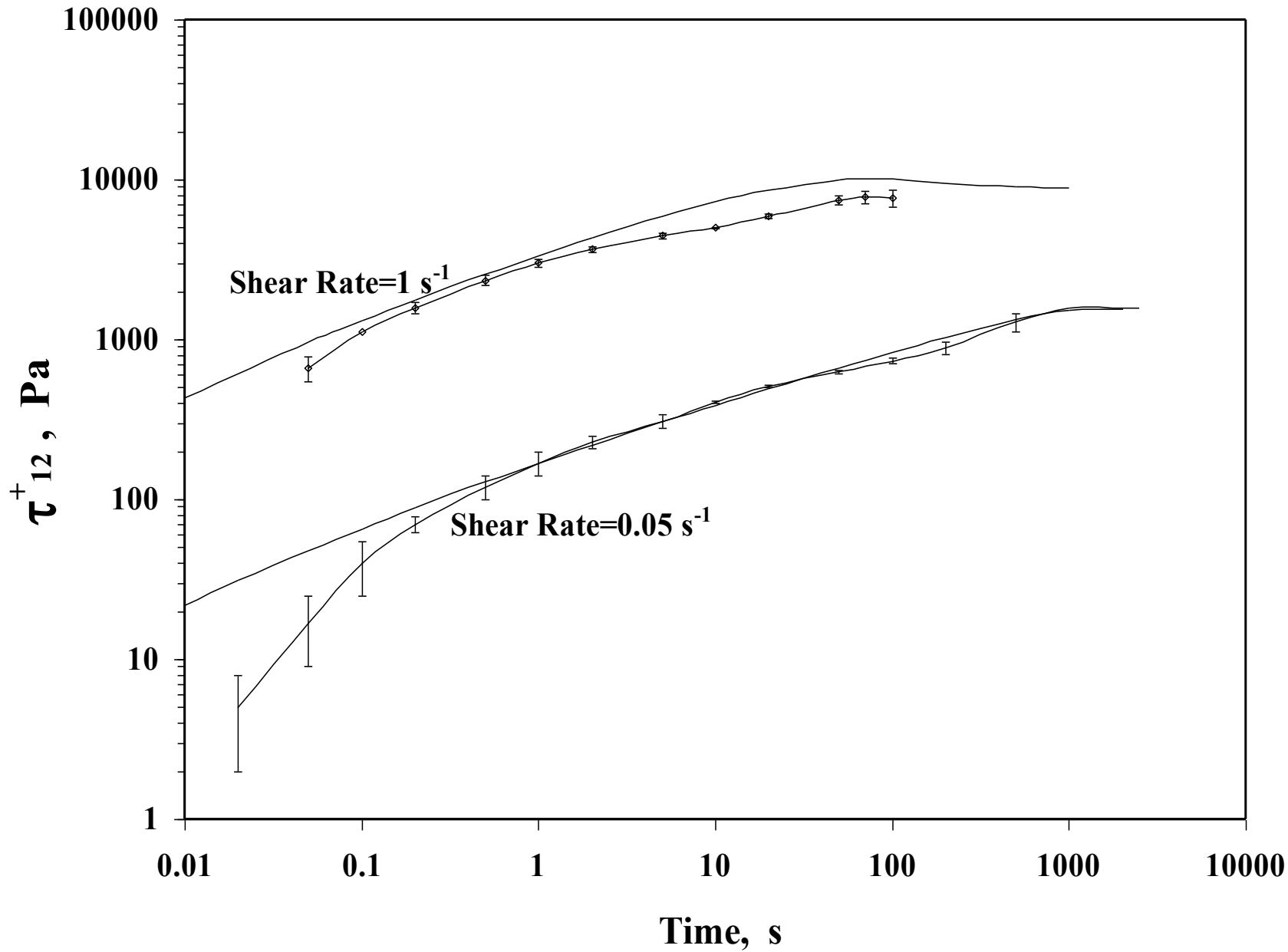

**Figure 9:** The experimental data and predictions (damping function obtained from first normal stress difference relaxation data) of the shear stress growth of the TPE at 100°C.

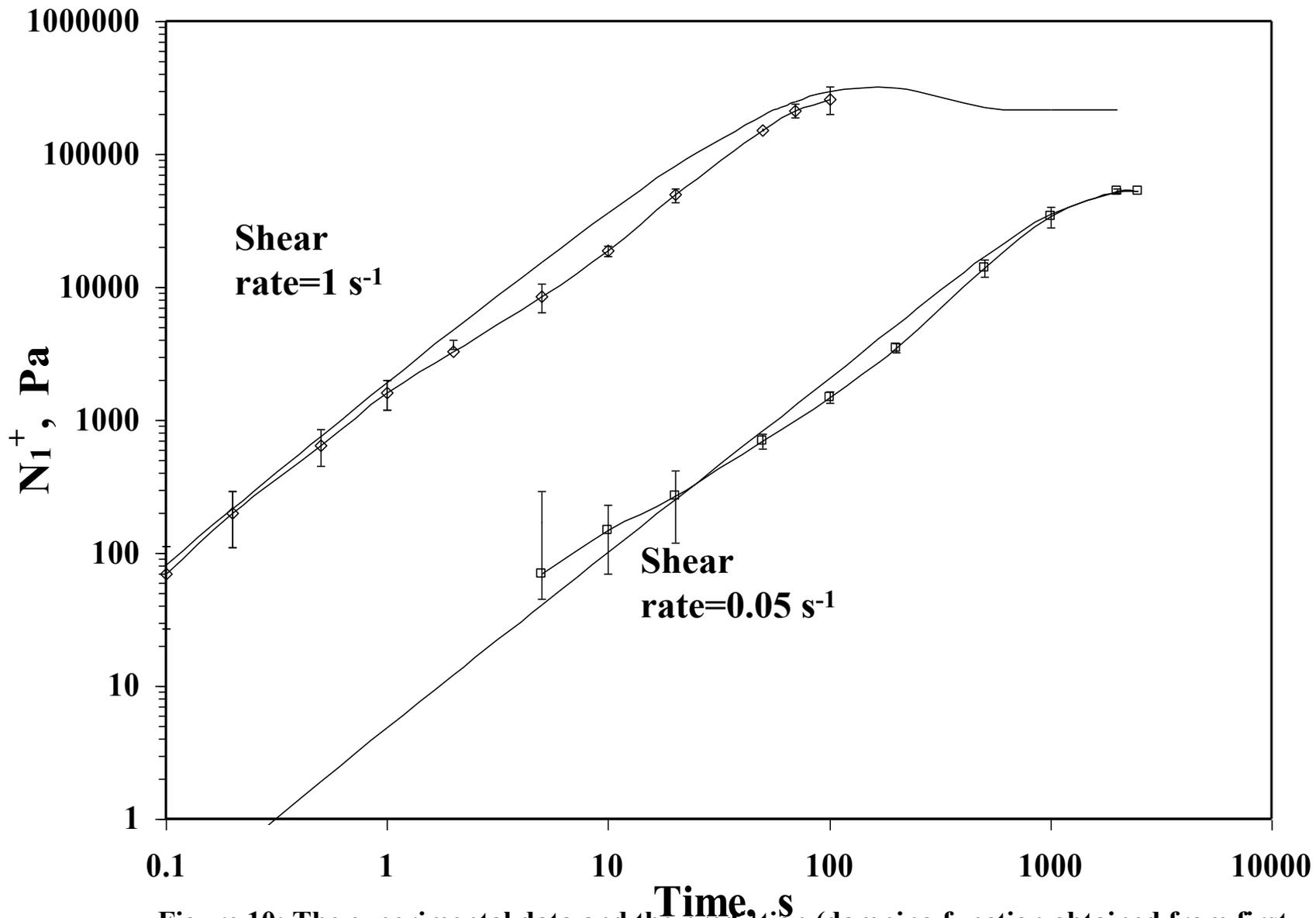

Figure 10: The experimental data and the prediction (damping function obtained from first normal stress difference relaxation data itself) of the first normal stress difference growth of the TPE at 100°C.

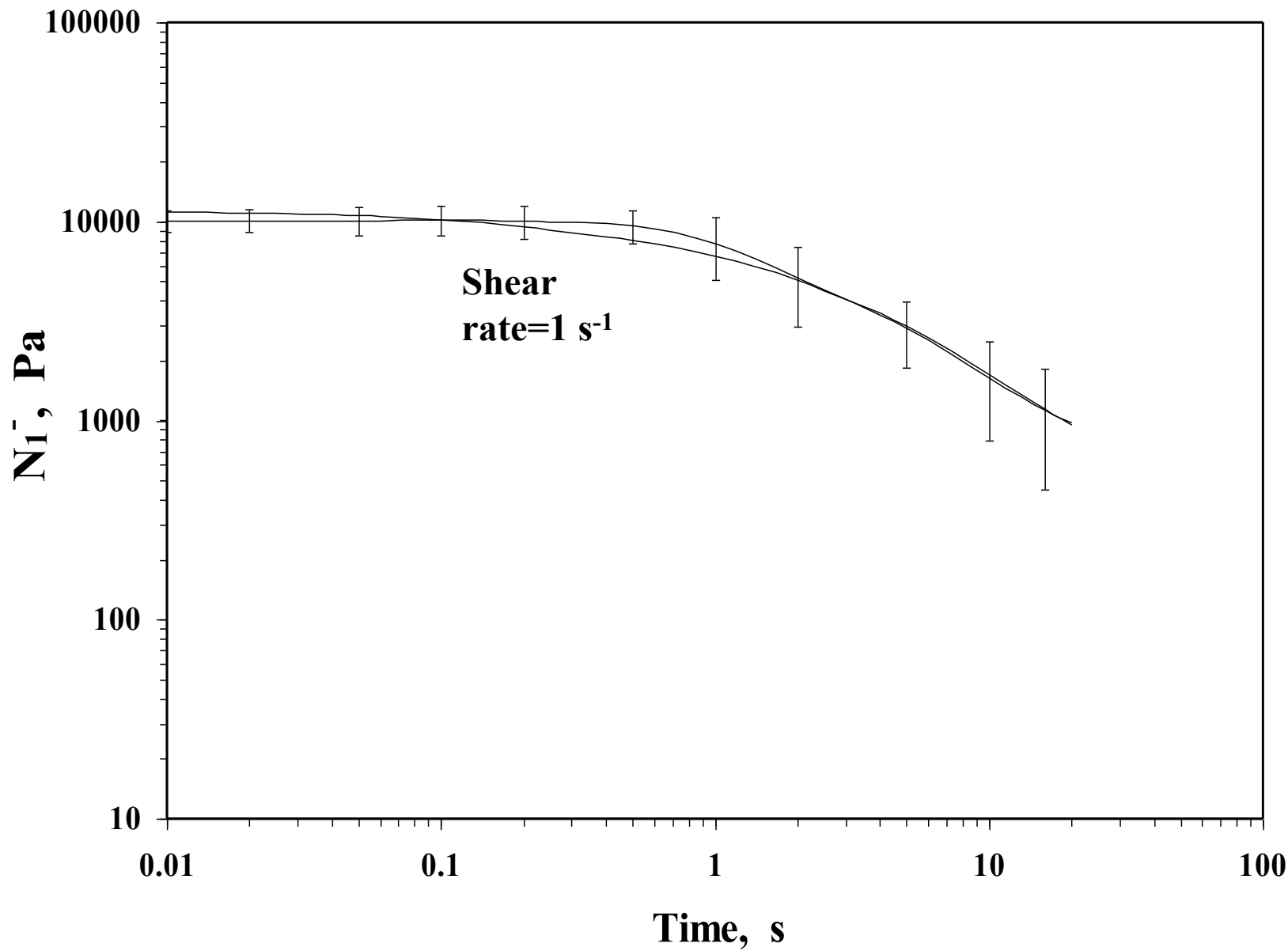

Figure 11: The experimental data and best fit (damping function obtained from first normal stress difference relaxation data) of the first normal stress difference relaxation of HDPE at 190°C.

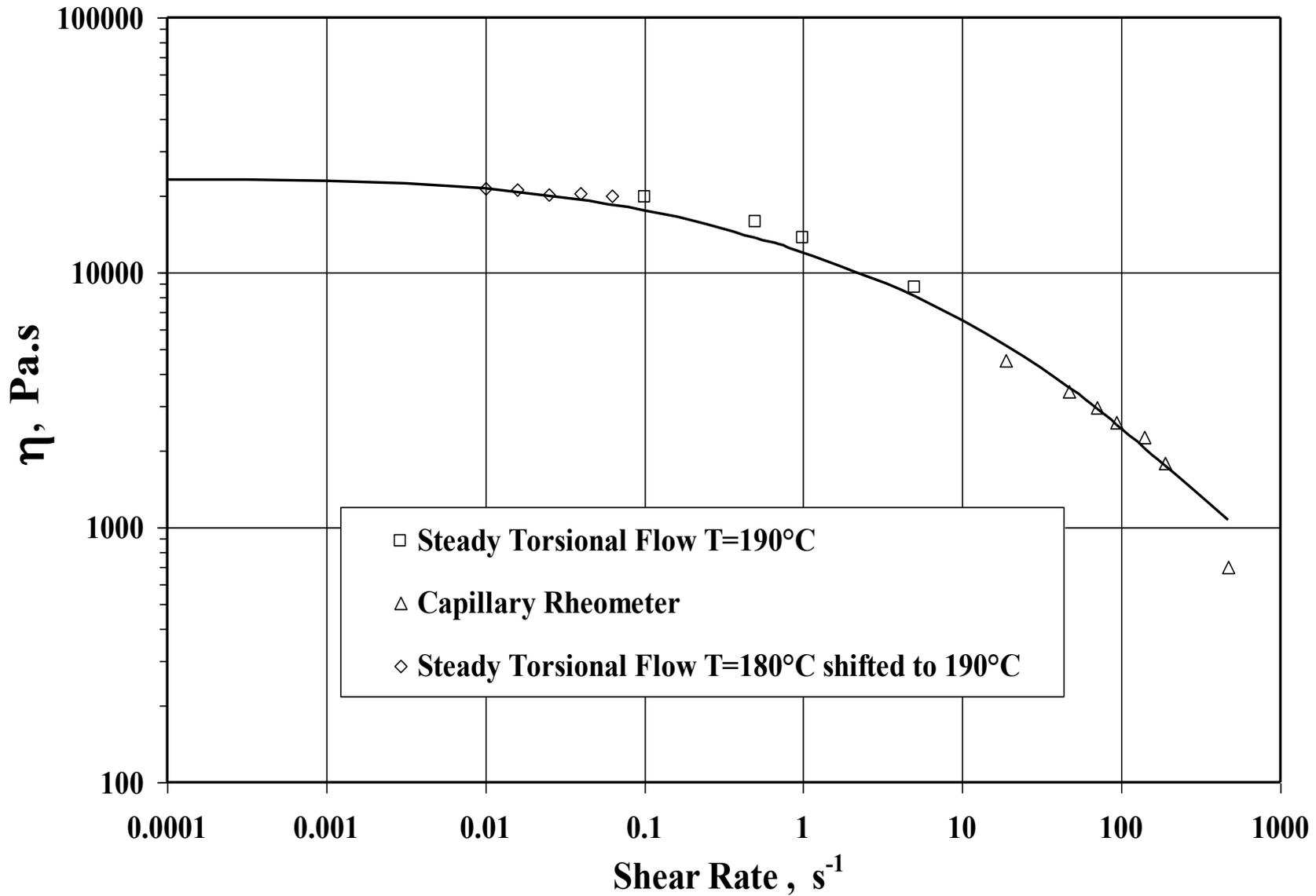

**Figure 12: The experimental data and predictions (damping function obtained from first normal stress difference relaxation data) of the shear viscosity of HDPE at 190°C.**

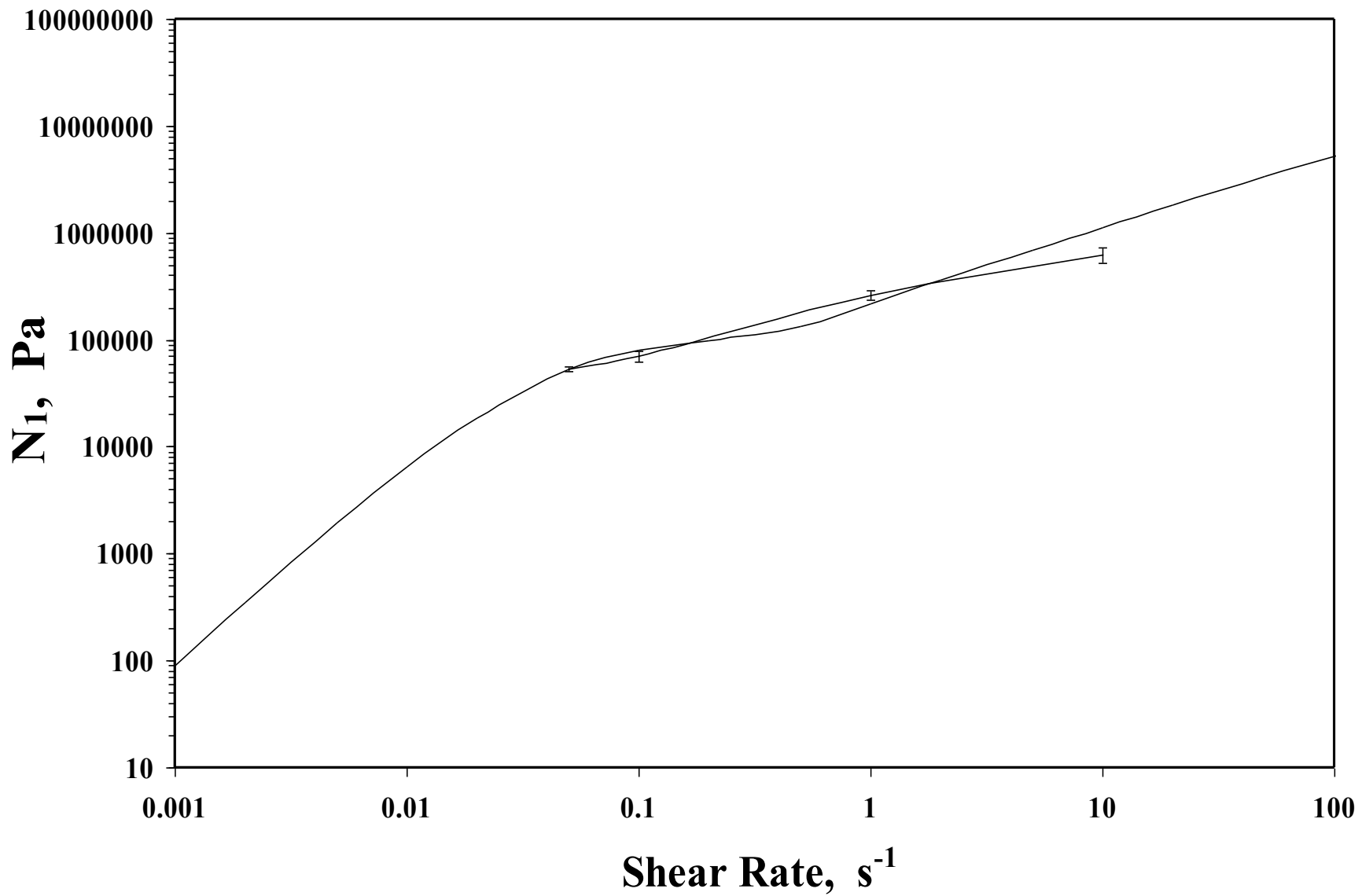

**Figure 13:** The experimental data and predictions (damping function obtained from first normal stress difference relaxation data) of the first normal stress difference of the TPE at 100°C.

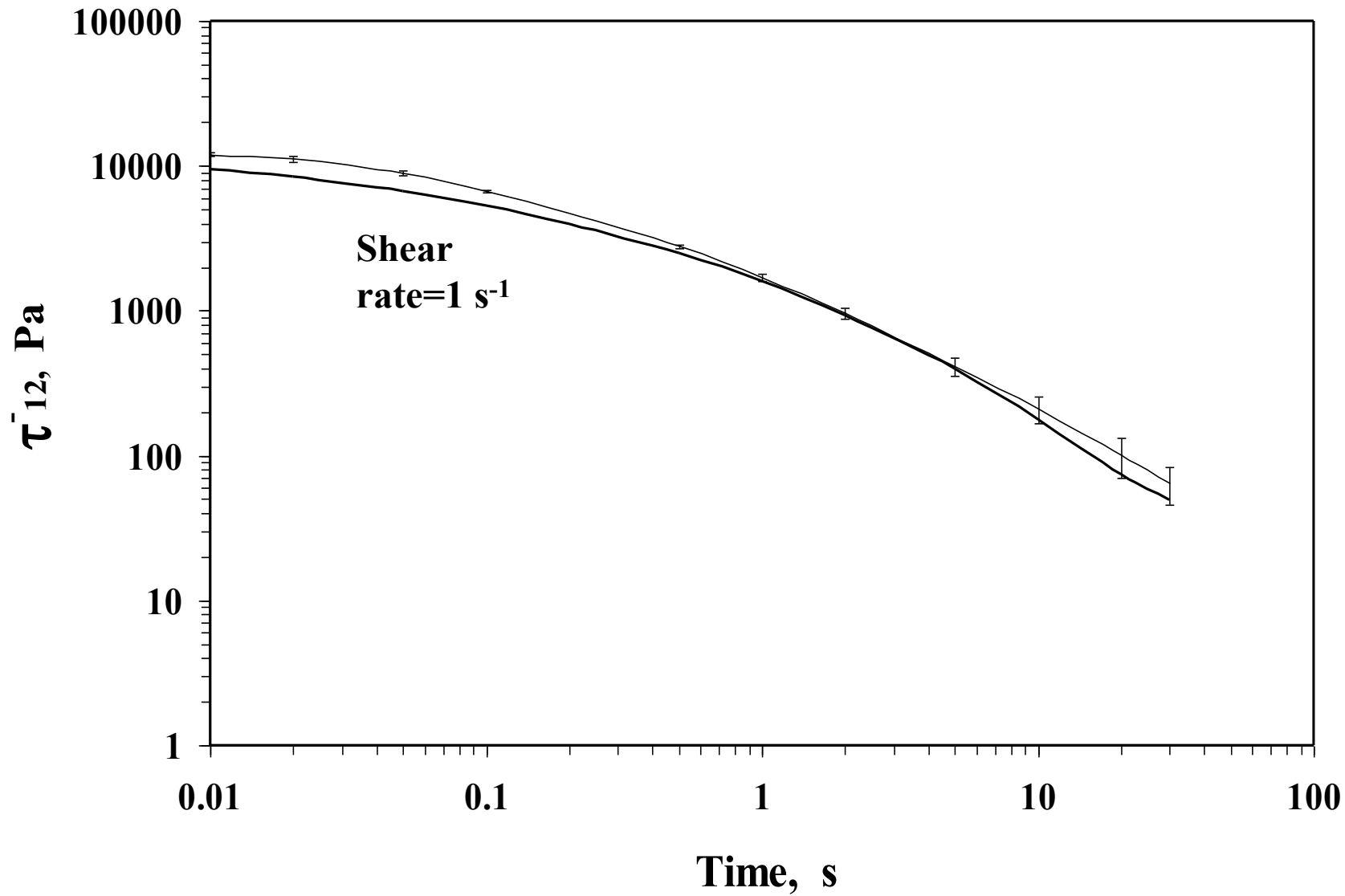

**Figure 14: The experimental data and predictions (damping function obtained from first normal stress difference relaxation data) of the shear stress relaxation of HDPE at 190°C.**

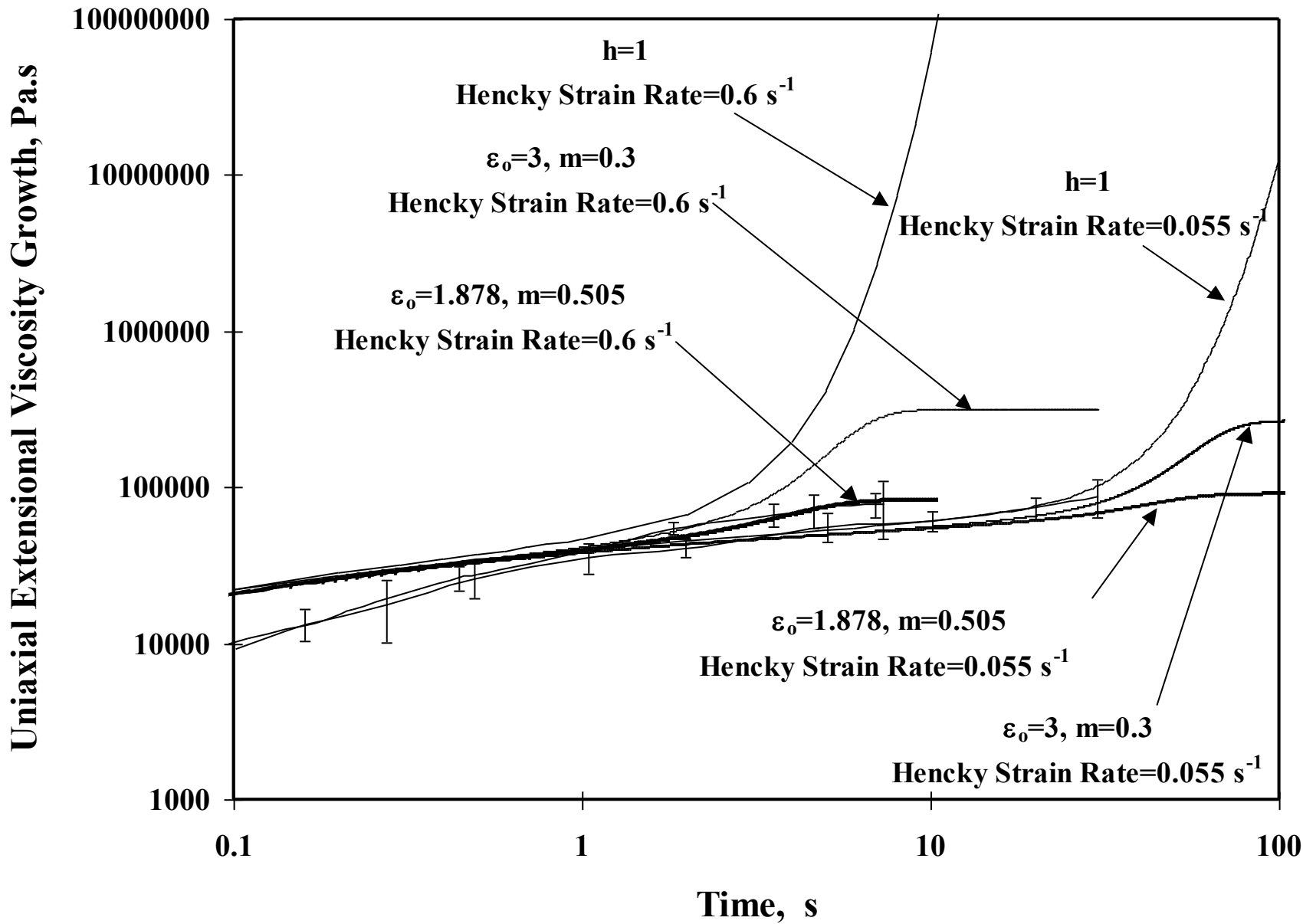

**Figure 15: The experimental data and predictions of the uniaxial extensional viscosity growth of HDPE at 190°C.**

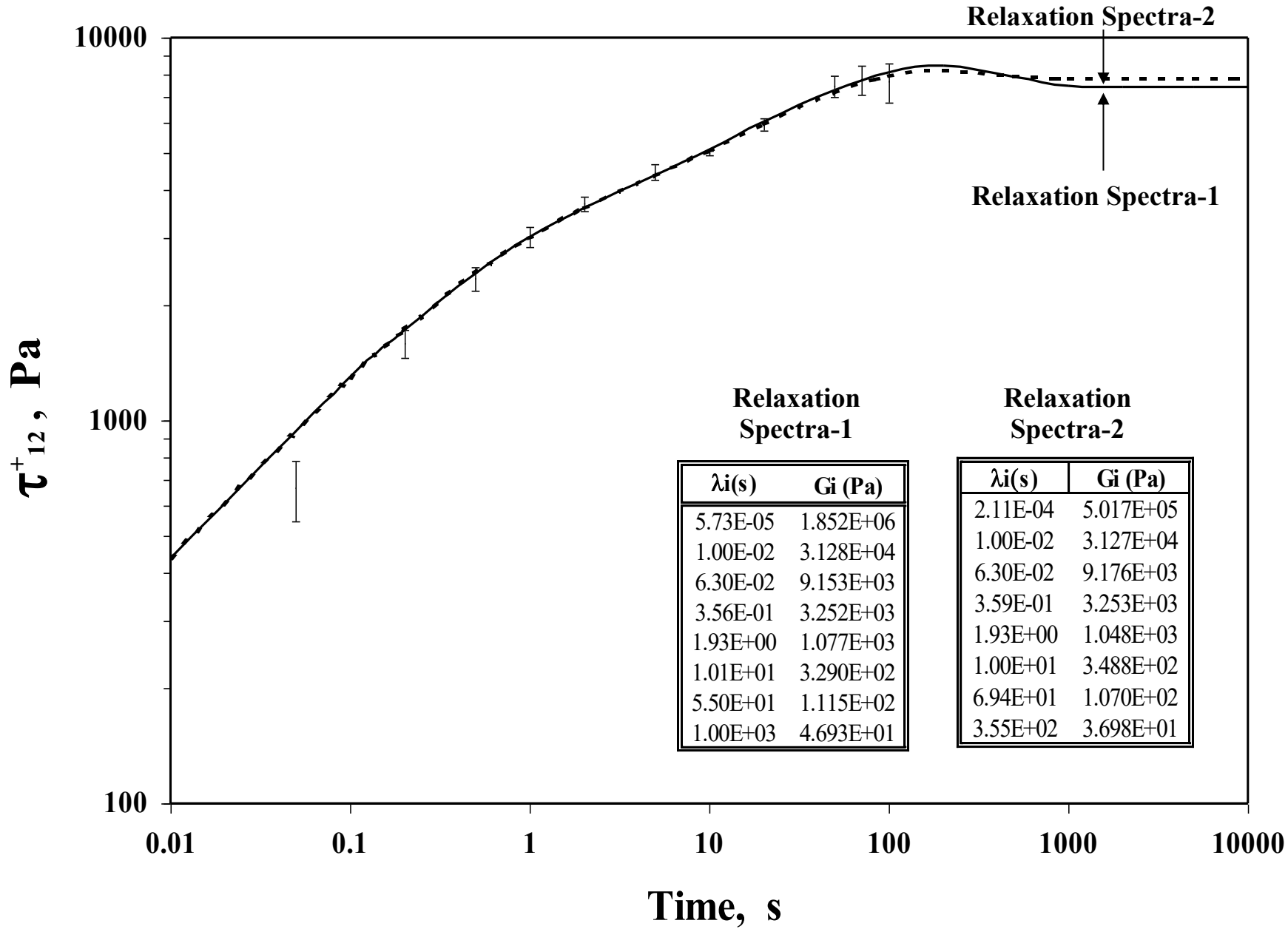

Figure 16: Predictions of the shear stress growth of TPE at 1 s$^{-1}$ using two different sets of relaxation spectra and similar damping function parameters of f=0.51, n$_1$=0.05, n$_2$=0.49.